\documentclass[12pt,letterpaper]{article}
\usepackage{graphicx}
\usepackage[sort&compress]{natbib}
\usepackage[usenames]{color}
\bibpunct{[}{]}{,}{a}{,}{,}
\newcommand{\noi}{\noindent}
\newcommand{\np}{\newpage\noindent}

\newcommand{\ns}{\normalsize}
\newcommand{\ls}{\noindent\large}
\newcommand{\Ls}{\noindent\Large}

\newcommand{\af}{\alpha}

\newcommand{\sg}{\sigma}

\newcommand{\Tb}{\bar{T}}
\newcommand{\1}{\bar{1}}
\newcommand{\2}{\bar{2}}
\newcommand{\3}{\bar{3}}
\newcommand{\4}{\bar{4}}
\newcommand{\5}{\bar{5}}
\newcommand{\6}{\bar{6}}
\newcommand{\7}{\bar{7}}

\begin{document}
\begin{center}
\vspace*{1in}
\Ls {\bf Matrix Representation of Knots and Folds: I}\ns
\vspace{1in}

                R. Kariotis   \\
         {\em Department of Physics \\
        University of Wisconsin \\
          Madison, Wisconsin 53706} \\
       (/knots/matrix.tex --  20feb12) \\
\end{center}

{\bf OVERVIEW:}
This report presents a unified treatment of the density of states for the
knots and folds  of polymer chains. The physical realization of such
systems ranges from DNA molecules (Taylor) to the microscopic  configurations
of space-time (Rovelli,Baez).
Explicit calculations employing this procedure will appear in a
subsequent paper.

The method in brief is as follows:\\
\noi
1. the starting point is a straight line/chain connecting two boundaries
A and B as shown in Fig 1\\
\noi
2. the line is bent at each node, starting  at one end and successively
moving down the  length of the chain one node at a time\\
\noi
3. at each point where the chain crosses itself a "cross-over/under" is
assigned\\
\noi
4. a complete set of configurations, the function space of the knot/fold,
is obtained by allowing all possible folds/crossings
(sequences of folds/crossings are characterized by binary strings)\\
\noi
5. a matrix construction is used to characterize the topologically distinct
configurations;  we comment on the implied curvature of the resulting space 
\\
\\
\\
\noindent
bobk@physics.wisc.edu
\np
\noi
{\ls\bf PART A: folds}

\noi
{\bf Motivation}

It has been said that the phenomenon of polymer folding is an intractable
physics problem, yet at first glance the problem appears to be well suited to
the machinery of statistical mechanics: first choose a model pair potential
between joints in the chain, then use the total energy in the Boltzmann
factor and sum over all possible configurations. Iteration of this procedure
then follows by suitable adjustments to the assumed pair potential.
A difficulty arises in that the pair potential is likely to be
dependent on the entire configuration; this is almost certainly due to
quantum mechanical influence on the electron orbitals that determine the
pair potential.
Failure of statistical mechanical models suggests the need to separate energy and
configurational considerations.
\\
\\
\noi
{\bf 1. Statistical Mechanics}\\

Models of the self-avoiding walk are frequently used to describe the
quantitative properties of polymer/protein chain folding
(Freed, Pande et al, Amit et al). Typically these tools
employ standard methods in statistical mechanics. A
multi-particle Hamiltonian is set up
\[
                 H=\sum_{i,j} V(x_{i},x_{j})
\]
representing a fictitious energy of interaction between
two elements in the chain located at $x_{i}$ and $x_{j}$ . The
statistical properties of the chain are then obtained from analysis
of the partition function
\[
                Z=\int D[x_{i}]e^{-\beta H}
\]
where $\beta$ acts as an inverse temperature parameter. It is
a straight forword procedure to evaluate this quantity, but because the
interactions must be strong  to effectively eliminate configurations
where the chain crosses itself, standard perturbative techniques are
of limited value.

Evaluation of the partition function is not trivial of course, but nevertheless
provides a complete and systematic solution to the problem. The objection to this approach is
that the effective/model potential is likely to be unrepresentative. The energy
of interaction between sites is unavoidably quantum mechanical and thus the
interaction will vary not just on distance between pairs but also on total
chain configuration (Bryngelson 1994).
Typically, simulations have found the need for what are called non-native
interactions (Wallin et al) in order to match with experiment.

To this extent we set energy considerations aside and limit the discussion to
configurational concerns.
Exact pair interaction must almost always depend on total configuration, hence, 
non-native interactions.
\\
\\

\np
\noi
{\bf 2. The Folding Operator}\\

Each initial chain is a line on the interval $[0,N]$. Each fold describes a bend of 
$\pm\frac{\pi}{2}$, the first bend being at the node $N-1$, the second at $N-2$, etc.
A  binary string $[00101...011]$ is used to charcterize the $\pm$ sequence for one complete fold,
while
a second is used to  describe the set of crossings as shown in Fig 1. The matrices
that perform this operations are as follows.

\begin{figure}[!h]
\begin{center}
\includegraphics[width=3in,height=3in,trim=.5in .5in .5in .5in,angle=90,scale=1]{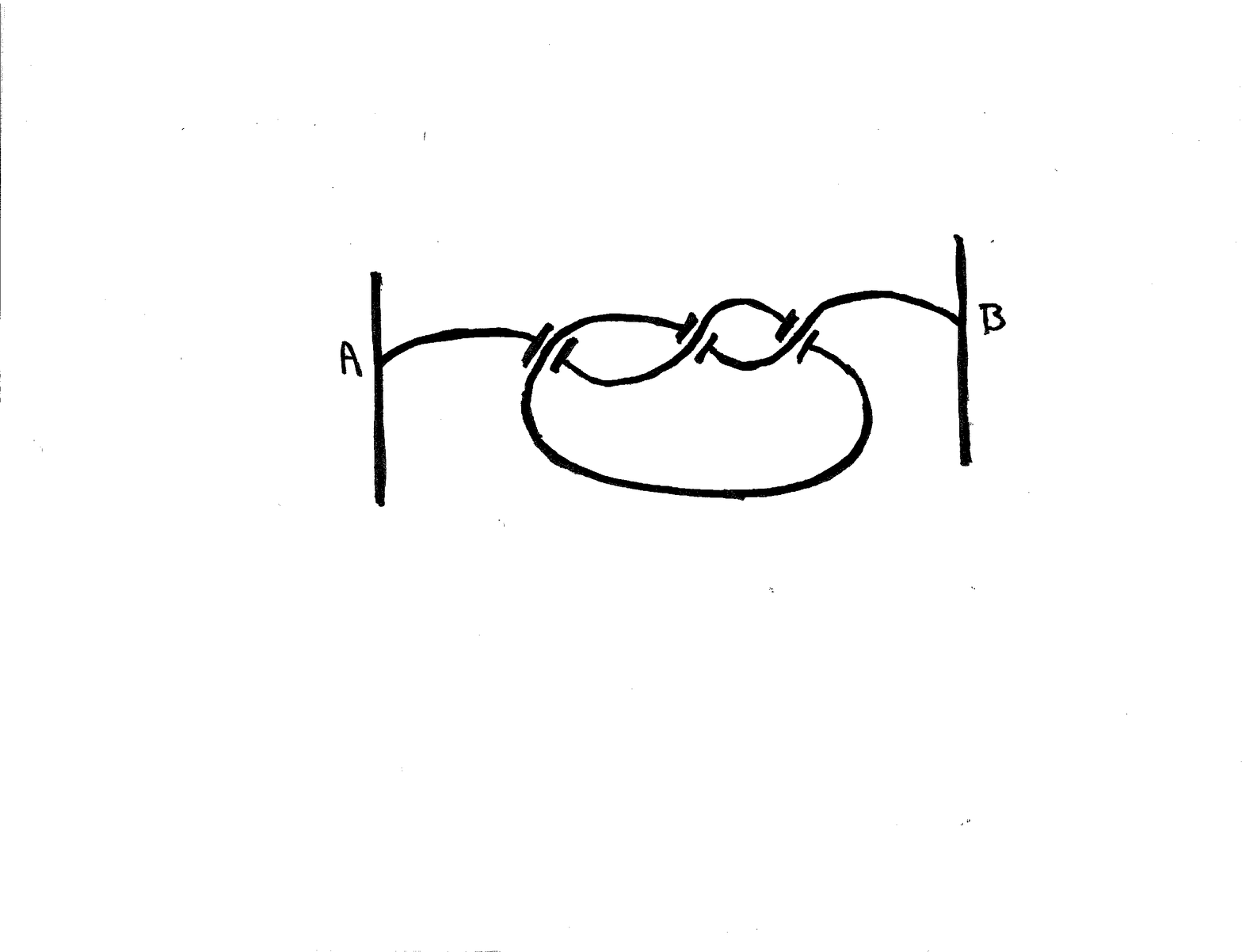}
\caption{typical configuration, in this case $3_{1}$, the trefoil }
\end{center}
\end{figure}

\begin{figure}[!h]
\begin{center}
\includegraphics[width=2.5in,height=2in]{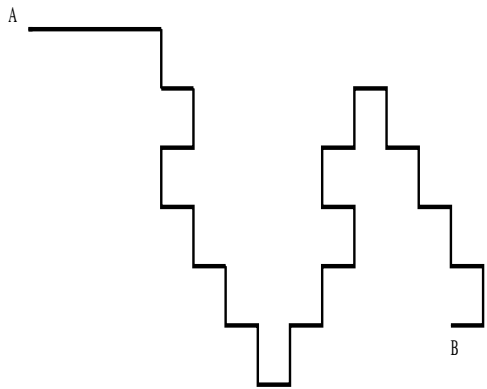}
\caption{typical bent polymer without crossings }
\end{center}
\end{figure}

The definition of the folding operator is that it bends, at right angles,
an initially straight chain, one site after another. There is one such
operator, or matrix in this formulation, for each configuration, 
ignoring self-intersections, and it will be the goal of the analysis to pick
out those matrices that create allowable configurations.

The folding operator is easy to construct and relatively easy to diagonalize
but we will find limits to the extent of useful information obtained.

In general the rotation operator takes the form
\[
\tau(\theta)=
\left[
\begin{array}{cc}
                    \cos(\theta) &  \sin(\theta) \\
                   -\sin(\theta) &  \cos(\theta) \\
\end{array}
\right]
\]
where $\theta$ is the clockwise angle of rotation from the positive x-axis
(3 o'clock position). For  $\theta=\frac{\pi}{2}$
\[
\tau(\frac{\pi}{2})=
\left[
\begin{array}{cc}
                   0 &  1 \\
                   -1 &   0 \\

\end{array}
\right] \equiv \tau
\]
%%%%%%%%%%%%%%%%%%%%%%%%%%%%%%%%%%%%%%%%%%%%%%%%%%%%%%%%%%%%%%%%%%%%%%%%%%%%%%%%%%%%%%%%%%%%%%%%%%%%%%%%%%%%%%%%
%%%%%%%%%%%%%%%%%%%%%%%%%%%%%%%%%%%%%%%%%%%%%%%%%%%%%%%%%%%%%%%%%%%%%%%%%%%%%%%%%%%%%%%%%%%%%%%%%%%%%%%%%%%%%%%%
%%%%%%%%%%%%    matrices
\newcommand{\idn}{
\left[
\begin{array}{cc}
                   1 &  0 \\
                   0 &  1 \\
\end{array}
\right]
                }
\newcommand{\tv}{
\left[
\begin{array}{cc}
                   0 &  1 \\
                   -1 &  0 \\
\end{array}
\right]
                }
\newcommand{\swt}{
\left[
\begin{array}{c  c }
              0 & 1 \\
              1 & 0 \\
\end{array}
\right]
}

\newcommand{\tva}{
\left[
\begin{array}{c  c }
              1 & -1  \\
              1 &  1 \\
\end{array}
\right]
}
\newcommand{\pd}{\ddots}
\newcommand{\tvb}{
\left[
\begin{array}{c  c }
              -1 &  1  \\
              -1 & -1 \\
\end{array}
\right]
}
\newcommand{\idN}{
\left[
\begin{array}{ccc}
              1  &       &   \\
                 &\ddots &  \\
                 &       & 1 \\ 
\end{array}
\right]
}

\newcommand{\pva}{
\left[
\begin{array}{c  c }
              1       &      \\
              \sg_{1} & \tau_{1} \\
\end{array}
\right]
}
\newcommand{\pvb}{
\left[
\begin{array}{c  c c}
                1     &           &  \\
              \sg_{2} & \tau_{2} &  \\
              \sg_{2} &          & \tau_{2}\\
\end{array}
\right]
}
\newcommand{\pvc}{
\left[
\begin{array}{c  c c c }
              1       &           &           &    \\
              \sg_{3} & \tau_{3}  &           &  \\
              \sg_{3} &           & \tau_{3}  &  \\
              \sg_{3} &           &           & \tau_{3} \\
\end{array}
\right]
}

%%%%%%%%%%%%%%%%%%%%%%%%%%%%%%%%%%%%%%%%%%%%%%%%%%%%%%%%%%%%%%%%%%%%%%%%%%%%%%%%%%%%%%%%%%%%%%%%%%%%%%%%%%%%%%%%%%%%
%%%%%%%%%%%%%%%%%%%%%%%%%%%%%%%%%%%%%%%%%%%%%%%%%%%%%%%%%%%%%%%%%%%%%%%%%%%%%%%%%%%%%%%%%%%%%%%%%%%%%%%%%%%%%%%%%%%%%

Define $ I_{n}=diag(1,1,...,1)$ n elements.
For a chain of n+1 sites set from (0,0) to (n,0), 
folding at the second to the last site is
\[ 
        I_{n-2} \oplus  \pva 
\]
where $\tau_{l}=s_{l}\tau$, $s=\pm 1$;
 folding at site third  from the end is
\[
          I_{n-3}\oplus  \pvb
\]
etc.

For example, successive applications of the first four
of these matrices yields
\[
\left[
\begin{array}{ccccc}
  1        &                 &                         &                     &   \\
  \sg_{4}  & \tau_{4}        &                         &                     &   \\
  \sg_{4}  & \tau_{4}\sg_{3} & \tau_{43}                &                    &   \\
  \sg_{4}  & \tau_{4}\sg_{3} & \tau_{43}\sg_{2}         &  \tau_{432}        &   \\
  \sg_{4}  & \tau_{4}\sg_{3} & \tau_{43}\sg_{2}          & \tau_{432}\sg_{1} & \tau_{4-1}   \\
\end{array}
\right]
\]
where
$\sg_{n} =1-\tau_{n}$, and
$\tau_{432}=s_{4}s_{3}s_{2}\tau^{3}$.

These are random matrices to be dealt with in the follow-up paper where
we discuss the distribution of eigenvalues. Preliminary numerical studies suggest
allowed structures tend to cluster in groups. Such behavior has been reported in
previous studies (Balafras and Dewey, Moret et al) and has possible application
to the so-called Levinthal Paradox (Karplus, Dill and Chan).

The Levinthal Paradox says that a large polymer would take eons rather than
milliseconds to fold if it did so by randomly sampling accessible states.
What this suggests is that the folding process (and knotting as well) is
accomplished through some globally determined energy potential and is not
a locally driven phenomenon (Wallin et al). 
\\
\\
\\
\\
%%%%%%%%%%%%%%%%%%%%%%%%%%%%%%%%%%%%%%%%%%%%%%%%%%%%%%%%%%%%%
\np
\noi
{\ls\bf Part B: Knots}\\

\noi
{\bf Motivation}

The difference between a knot and a fold in this  formulation of
the problem is the nature of the sequence of crossings:
a particular sequence of crossings may, or may not, actually
tie the chain, such that it becomes possible, or not, to
pull the two ends arbitrarily far apart, returning the
chain to the initial straight line.

How to do this is dealt in part by the following scheme.
Each knot is thought to be described by two dimensions,
position (or crossing) and time (or path length). 
The distinction as suggested in Fig 3
is made by first labeling all crossings successively until
each site has a number, followed by assigning to each site
a "time" value as a full circuit of the knot is traversed.

[Aside: For the first few knots all crossings are labeled before the
return path is begun; these are called sequential. However,
beginning with $7_{7}$ a previously labeled site is encountered
before all seven sites have been labeled; these are called
nonsequential. This is something like Euler's Konigsberg Bridge
problem. This distinction is made clear in Figs 4 and 5 where
we show how to reduce $6_{3}$ to a simpler circuit, thus making
it clear that the knot is sequential, and  also show that $7_{7}$
is necessarily nonsequential.]
\\
\\
\noi
{\bf 1. Knot Matrix Designation}

For a knot with N crossings:

\noi
a) each crossing is an element in a sequence of positions\\
b) the $n^{th}$ position element is described by the matrix
\[
T_{n}(t)=
        \left[
\begin{array}{ccc}
                       I_{n-1}  &                &                  \\
                                &  \tau(t)       &                  \\
                                &                &  I_{N-n-1}       \\
\end{array}
\right]           
 \]
where
\[
\tau(t) =
        \left[
\begin{array}{ccc}
                        0  &   e^{i\af t}    \\
                        1  &   0            \\
\end{array}
\right]           
 \]
for cross-over and $\tau^{\dagger}$ for cross-under; $\af$ to be determined.\\
c) the time sequence for a given knot is the product $[T]=T_{a}T_{b}T_{c}T_{d}...$
where
$abcd...$ are successive crossings visited on a complete path;
for example
for $6_{1}$ the $[T]$ product is
$\Tb_{5}T_{6}\Tb_{1}T_{2}\Tb_{3}T_{4}\Tb_{6}T_{5}\Tb_{4}T_{3}\Tb_{2}T_{1}$
abbreviated to $[\bar{5}6\bar{1}2\bar{3}4.\bar{6}5\bar{4}3\bar{2}1]$.
This procedure is not unlike the group theory method of Artin/Burau (Kauffman, p. 86)
except that
the operators are 2d and label crossings instead of braids (Birman and Brendle).
Also it appears that these matrices do not obey the Yang-Baxter relations
(Baez and Muniain).\\
\noi
d) as in Fig 3 each operator is 2d, i.e. $T_{a}(n)$ where
{\em (a,n)=(space,time)=(position,path length)}.

The matrices $\tau,\tau^{\dagger}$ act at each crossing something like raising and lowering
operators in field theory, and create a "time evolution" of the initial, 
ground state (the identity matrix),
carrying it through a sequence of intermediate states. The resulting matrix is a
measure of the curvature of the knot to the extent that the configuration is not
returned to the identity matrix for the closed contour.

\begin{figure}[!h]
\begin{center}
\includegraphics[width=3in,height=2in]{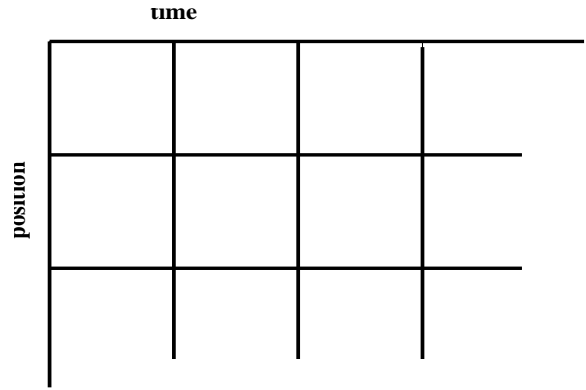}
\caption{ geometry of the matrix $T_{l}(n)$}
\end{center}
\end{figure}

\begin{figure}[!h]
\begin{center}
\includegraphics[width=5in,height=4in,trim=.5in .5in 1.5in .5in,clip,scale=1.1]{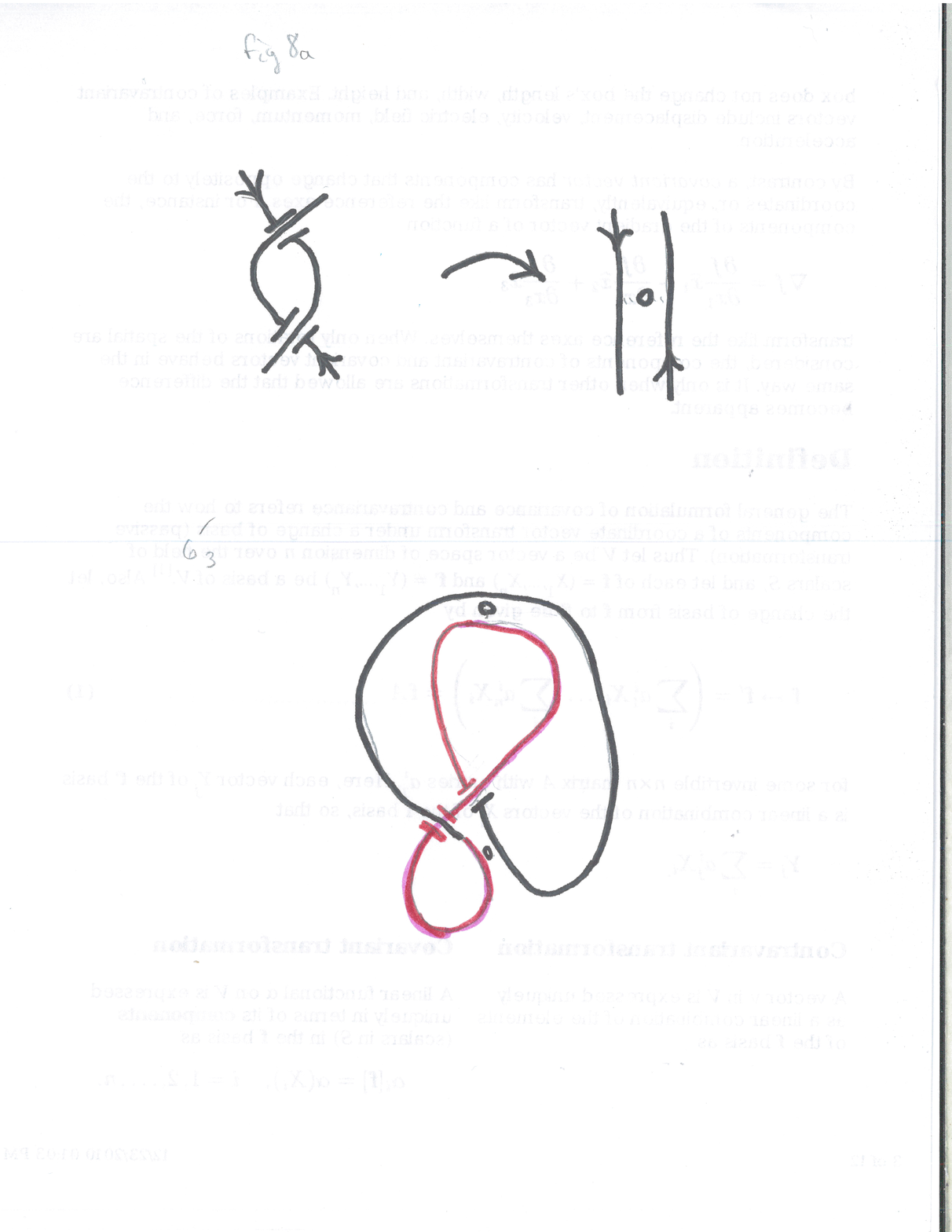}
\caption{ construction showing that $6_{3}$ is sequential}
\end{center}
\end{figure}

\begin{figure}[!h]
\begin{center}
\includegraphics[width=5in,height=4in,trim=.5in .5in 1.5in .5in,clip,scale=1.1]{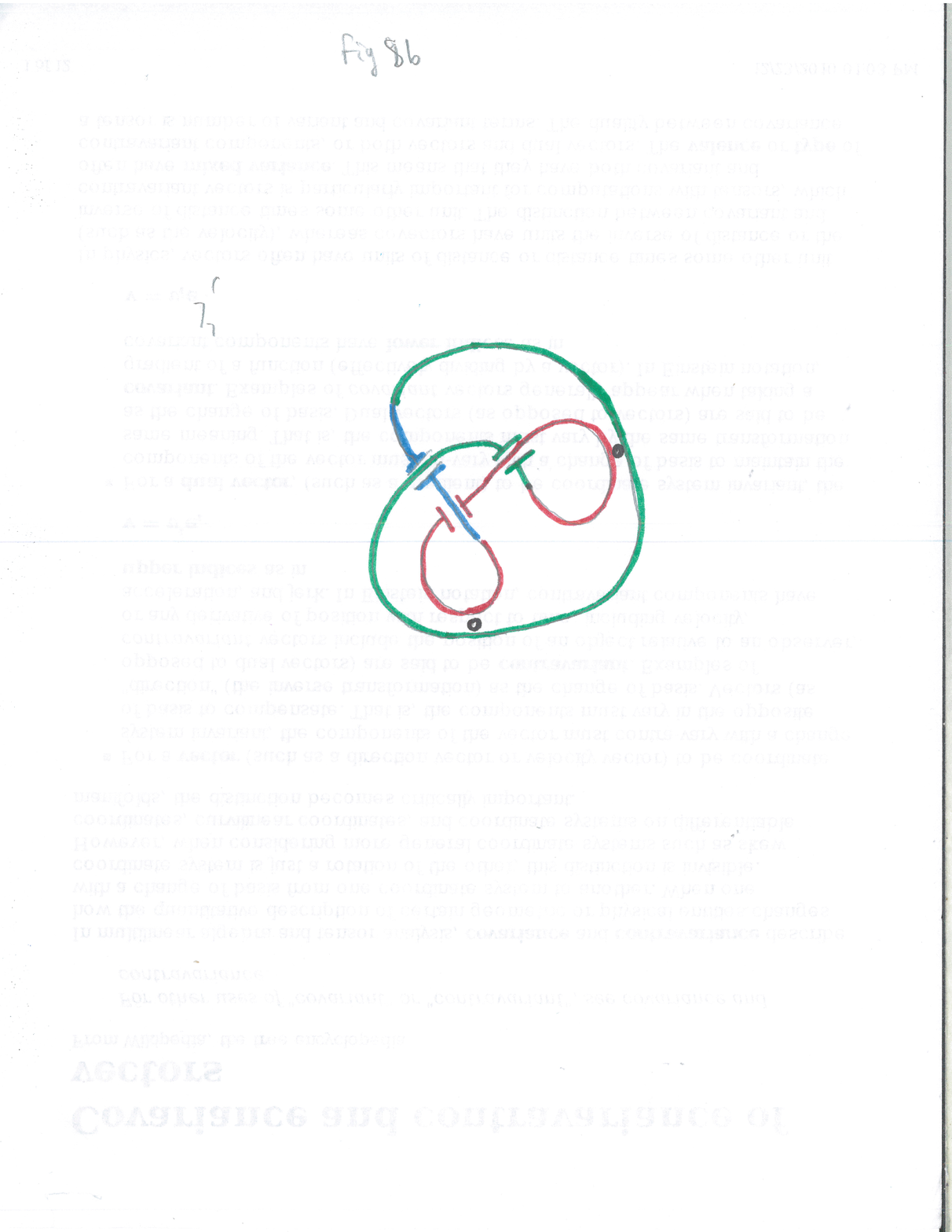}
\caption{ construction showing that $7_{7}$ is not sequential}
\end{center}
\end{figure}

\begin{figure}[!h]
\begin{center}
\includegraphics[width=3in,height=2in]{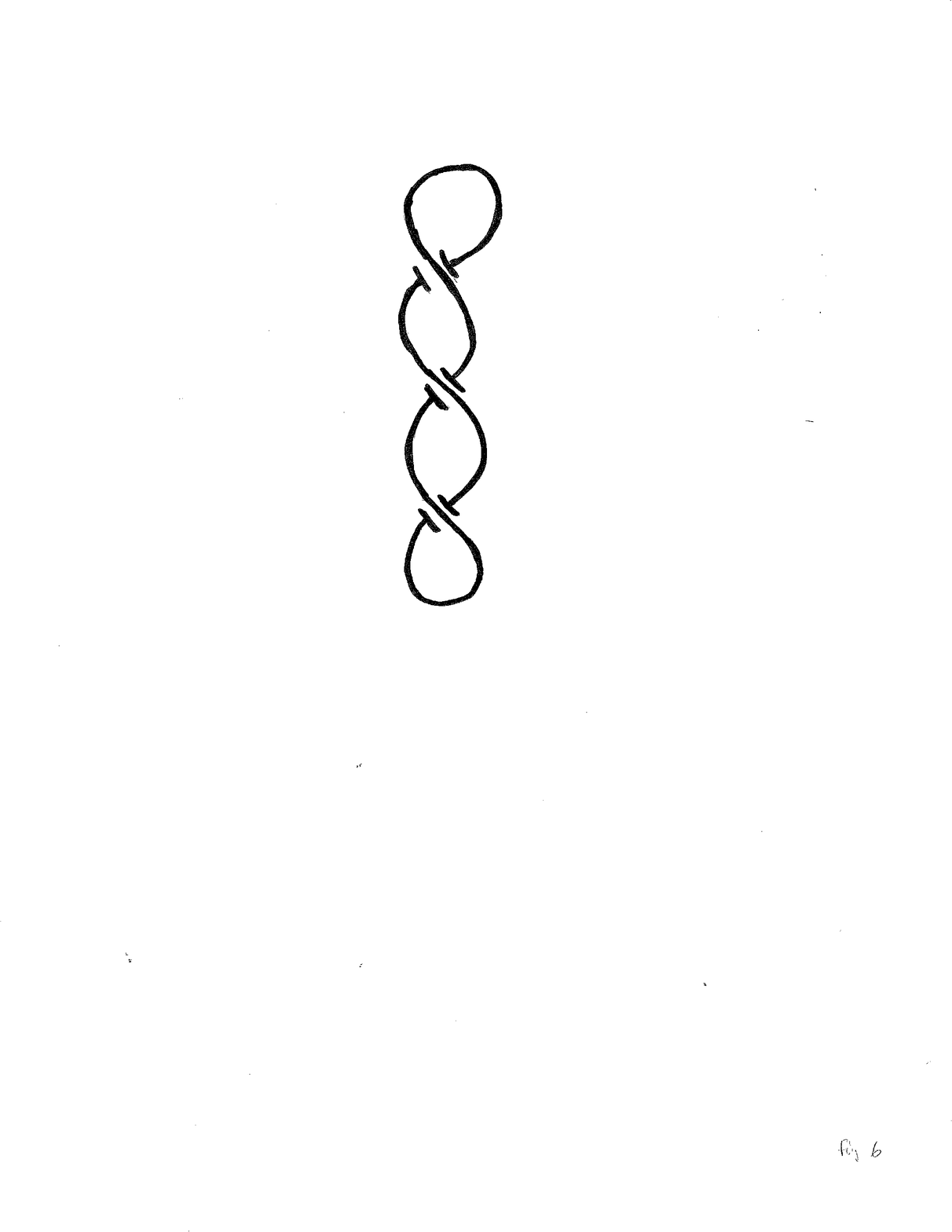}
\caption{ the identity operation for N=3}
\end{center}
\end{figure}

\np
{\bf 2. Topological Subspaces; Curvature}

In order to construct the representative matrices for each knot we need to
assign a value to the time variable $e^{i\af}$ and this is accomplished
by requiring that configurations isomorphic to the unknot give the
identity as shown in Fig 6. For numerical purposes we can choose
an arbitrary real number $q$ and require that the second pass
through the crossings act in reverse time; for example for $4_{1}$
the time factors are $q,q^{2},q^{3},q^{4},q^{4},q^{3},q^{2},q$.

The resulting matrices act as a representation of the curvature of the
knot in the sense that a complete path around the knot is analogous
to the closed curve in the computation of parallel displacement of
a vector defined on a smooth manifold. The resulting matrix bares
some relation to the curvature tensor in differential geometry;
in quantum field theory, this quantity is a relation of the Wilson Loop,
without the extra machinery for field theory purposes.

The table below gives the knot, the path and the resulting matrices for
the figures shown in the Appendix.
\\
\\
\noi
[Aside: I gave up trying to get LATEX to put the figures where I wanted.
The Appendix included here contains the relevant knots of the table;
the drawings, inexpert as they are, indicate the sequence of crossings
used to obtain the matrices below.]

%%%%%%%%%%%%%%%%%%%%%%%%%%%%%%%%%%%%%%%%%%%%%%%%%%%%%%%%%%%%%%%%%%%%%%%%%%%%%%%%%%%%%%%
%%%%%%%%%%%%%%%%%%%%%%%%%%%%%%%%%%%%%%%%%%%%%%%%%%%%%%%%%%%%%%%%%%%%%%%%%%%%%%%%%%%%%%
\[
\begin{array}{c c c}
%%%%%%%%%%%%%%%%%%%%%%%%%%%%%%%%%%%%%%%%%%%%%%%%%%%%%%%%%%%%%%%%%%%%%%%%%%%%%%%%%%%%%%
3_{1} & [\32\1.3\21] &
\left[
\begin{array}{cc|cc}
                           &        &  1  &    \\
                           &        &     & q^{-5} \\ \hline
                    q^{2}  &        &     &    \\
                           & q^{3}  &     &   \\
\end{array}
\right]
\\ \hline
%%%%%%%%%%%%%%%%%%%%%%%%%%%%%%%%%%%%%%%%%%%%%%%%%%%%%%%%%%%%%%%%%%%%%%%%%%%%%%%%%%%%%%
%%%%%%%%%%%%%%%%%%%%%%%%%%%%%%%%%%%%%%%%%%%%%%%%%%%%%%%%%%%%%%%%%%%%%%%%%%%%%%%%%%%%%%
4_{1} & [\34\12.\43\21] &
\left[
\begin{array}{ccc|cc}
                            &        &   & q^{-7} &  \\
                            &  q^{2} &   &    &  \\
                     q^{4}  &        &   &    &   \\ \hline
                            &        & 1 &    &  \\
                            &        &   &    & 1 \\
\end{array}
\right]
\\ \hline

%%%%%%%%%%%%%%%%%%%%%%%%%%%%%%%%%%%%%%%%%%%%%%%%%%%%%%%%%%%%%%%%%%%%%%%%%%%%%%%%%%%%%
%%%%%%%%%%%%%%%%%%%%%%%%%%%%%%%%%%%%%%%%%%%%%%%%%%%%%%%%%%%%%%%%%%%%%%%%%%%%%%%%%%%%%%
5_{1} & [\54\32\1.5\43\21] &
\left[
\begin{array}{ccc|ccc}
                            &        & 1 &     &   &   \\
                            &        &   & q^{-7}  &   &   \\
                            &        &   &     & 1 &   \\ \hline
                            &        &   &     &   & q^{-7}\\
                    q^{6}   &        &   &     &   &   \\
                            &   q^{5}&   &     &   &   \\
\end{array}
\right]
\\ \hline
%%%%%%%%%%%%%%%%%%%%%%%%%%%%%%%%%%%%%%%%%%%%%%%%%%%%%%%%%%%%%%%%%%%%%%%%%%%%%%%%%%%%%%
5_{2} & [\54\12\3.5\43\21] &
\left[
\begin{array}{ccc|ccc}
                            &        &    &    & q^{-4} &   \\
                            &    q^{2}   &    &    &    &   \\
                            &        &  1 &    &    &   \\ \hline
                            &        &    &    &    & q^{-7}\\
                        q^{6}   &        &    &    &    &   \\
                            &        &    & q^{3}  &    &   \\
\end{array}
\right]
\\ \hline
%%%%%%%%%%%%%%%%%%%%%%%%%%%%%%%%%%%%%%%%%%%%%%%%%%%%%%%%%%%%%%%%%%%%%%%%%%%%%%%%%%%%%%%
\end{array} \]
%%%%%%%%%%%%%%%%%%%%%%%%%%%%%%%%%%%%%%%%%%%%%%%%%%%%%%%%%%%%%%%%%%%%%%%%%%%%%%%%%%%%%%%%
\[\begin{array}{c c c}

6_{1} & [\56\12\34.\65\43\21] &
\left[
\begin{array}{cccc|ccc}
                            &        &   &   &   & q^{-15} &   \\
                            &   q^{2}&   &   &   &     &   \\
                            &        & 1 &   &   &     &   \\
                            &        &   & q^{2}&&     &   \\ \hline
                     q^{10} &        &   &   &   &     &   \\
                            &        &   &   & 1 &     &   \\
                            &        &   &   &   &     & 1 \\
\end{array}
\right]
\\ \hline
%%%%%%%%%%%%%%%%%%%%%%%%%%%%%%%%%%%%%%%%%%%%%%%%%%%%%%%%%%%%%%%%%%%%%%%%%%%%%%%%%%%%%%
6_{2} & [\54\36\12.\65\43\21] &
\left[
\begin{array}{cccc|ccc}
                            &        &   & q^{-9}   &    &    &   \\
                            & q^{4}  &   &      &    &    &   \\
                            &        &   &      &  1 &    &   \\
                            &        &   &      &    & q^{-7} &   \\ \hline
                     q^{8}  &        &   &      &    &    &   \\
                            &        & q^{4} &      &    &    &   \\
                            &        &   &      &    &    & 1 \\
\end{array}
\right]
\\ \hline
%%%%%%%%%%%%%%%%%%%%%%%%%%%%%%%%%%%%%%%%%%%%%%%%%%%%%%%%%%%%%%%%%%%%%%%%%%%%%%%%%%%%%%

6_{3} & [\36\52\14.\65\43\21] &
\left[
\begin{array}{cccc|ccc}
                            &        & 1 &     &   &     &   \\
                            &        &   & q^{-7}  &   &     &   \\
                            &        &   &     &   & q^{-11} &   \\
                            &        &   &     &   &     & 1 \\ \hline
                            &    q^{5}   &   &     &   &     &   \\
                  q^{10}    &        &   &     &   &     &   \\
                            &        &   &     & q^{3} &     &   \\
\end{array}
\right]
\\ \hline

%%%%%%%%%%%%%%%%%%%%%%%%%%%%%%%%%%%%%%%%%%%%%%%%%%%%%%%%%%%%%%%%%%%%%%%%%%%%%%%%%%%%%%%
%%%%%%%%%%%%%%%%%%%%%%%%%%%%%%%%%%%%%%%%%%%%%%%%%%%%%%%%%%%%%%%%%%%%%%%%%%%%%%%%%%%%%%

7_{1} & [\76\54\32\1.7\65\43\21] &
\left[
\begin{array}{cccc|cccc}
                            &          &   1   &       &       &      &   & \\
                            &          &       & q^{-9}   &       &      &   & \\
                            &          &       &        &   1   &      &   & \\
                            &          &       &      &       &  q^{-9}  &   & \\ \hline
                            &          &       &      &       &      & 1 & \\
                            &          &       &      &       &      &   & q^{-9}\\
                     q^{12} &          &       &      &       &      &  & \\
                            &   q^{15} &       &      &       &      &   & \\
\end{array}
\right]
\\  \hline
%%%%%%%%%%%%%%%%%%%%%%%%%%%%%%%%%%%%%%%%%%%%%%%%%%%%%%%%%%%%%%%%%%%%%%%%%%%%%%%%%%%%%%
\end{array} \]
%%%%%%%%%%%%%%%%%%%%%%%%%%%%%%%%%%%%%%%%%%%%%%%%%%%%%%%%%%%%%%%%%%%%%%%%%%%%%%%%%%%%%%
\[\begin{array}{c c c}

7_{2} & [\76\12\34\5.7\65\43\21] &
\left[
\begin{array}{cccc|cccc}
                            &          &       &      &       &      & q^{-10} & \\
                            &    q^{2}     &       &      &       &      &  & \\
                            &          &   1   &      &       &      &  & \\
                            &          &       &  q^{2}   &       &      &  & \\ \hline
                            &          &       &      &  1    &      &  & \\
                            &          &       &      &       &      &  & q^{-9}\\
                    q^{12}  &          &       &      &       &      &  & \\
                            &          &       &      &       &  q^{3}   &  & \\
\end{array}
\right]
\\ \hline
%%%%%%%%%%%%%%%%%%%%%%%%%%%%%%%%%%%%%%%%%%%%%%%%%%%%%%%%%%%%%%%%%%%%%%%%%%%%%%%%%%%%%%%%
7_{3} & [\56\74\32\1.7\65\43\21] &
\left[
\begin{array}{cccc|cccc}
                            &          &    1  &      &       &      &   & \\
                            &          &       &  q^{-9}  &       &      &   & \\
                            &          &       &      &   1   &      &   & \\
                            &          &       &      &       &  q^{-9}  &   & \\ \hline
                        q^{10}  &          &       &      &       &      &   & \\
                            &     q^{12}   &       &      &       &      &  & \\
                            &          &       &      &       &      & 1  & \\
                            &          &       &      &       &      &   & q^{-4}\\
\end{array} 
\right]
\\ \hline

%%%%%%%%%%%%%%%%%%%%%%%%%%%%%%%%%%%%%%%%%%%%%%%%%%%%%%%%%%%%%%%%%%%%%%%%%%%%%%%%%%%%%%

7_{4} & [\56\74\12\3.7\65\43\21] &
\left[
\begin{array}{cccc|cccc}
                            &          &       &      &   q^{-6}  &      &   & \\
                            &     q^{4}    &       &      &       &      &   & \\
                            &          &   1   &      &       &      &   & \\
                            &          &       &      &       &  q^{-9}  &   & \\ \hline
                         q^{10} &          &       &      &       &      &   & \\
                            &          &       &  q^{5}   &       &      &   & \\
                            &          &       &      &       &      & 1 & \\
                            &          &       &      &       &      &   & -q^{4}\\
\end{array}
\right]
\\
%%%%%%%%%%%%%%%%%%%%%%%%%%%%%%%%%%%%%%%%%%%%%%%%%%%%%%%%%%%%%%%%%%%%%%%%%%%%%%%%%%%%%%%%%
\end{array}\]
%%%%%%%%%%%%%%%%%%%%%%%%%%%%%%%%%%%%%%%%%%%%%%%%%%%%%%%%%%%%%%%%%%%%%%%%%%%%%%%%
\[\begin{array}{c c c}

7_{5} & [\76\32\14\5.7\65\43\21] &
\left[
\begin{array}{cccc|cccc}
                            &          &   1   &      &       &      &   & \\
                            &          &       &  q^{-7}  &       &      &   & \\
                            &          &       &      &       &      & q^{-6}  & \\
                            &   q^{7}      &       &      &       &      &   & \\ \hline
                            &          &       &      &  1    &      &   & \\
                            &          &       &      &       &      &   & q^{-9}\\
                        q^{12}  &          &       &      &       &      &   & \\
                            &          &       &      &       &   q^{3}  &   & \\
\end{array}
\right]
\\ \hline

%%%%%%%%%%%%%%%%%%%%%%%%%%%%%%%%%%%%%%%%%%%%%%%%%%%%%%%%%%%%%%%%%%%%%%%%%%%%%%%%%%%%%%

7_{6} & [\56\12\74\3.7\65\43\21] &
\left[
\begin{array}{cccc|cccc}
                            &          &       &      &  q^{-4}   &      &   & \\
                            &    q^{2}     &       &      &       &      &   & \\
                            &          &    1  &      &       &      &   & \\
                            &          &       &      &       & q^{-11}  &   & \\ \hline
                         q^{10} &          &       &      &       &      &   & \\
                            &          &       &  q^{5}   &       &      &   & \\
                            &          &       &      &       &      & 1 & \\
                            &          &       &      &       &      &   & q^{-2}\\
\end{array}
\right]
\\ \hline

%%%%%%%%%%%%%%%%%%%%%%%%%%%%%%%%%%%%%%%%%%%%%%%%%%%%%%%%%%%%%%%%%%%%%%%%%%%%%%%%%%%%%%

7_{7} & [\57\34\76\1.7\65\43\21] &
\left[
\begin{array}{cccc|cccc}
                            &          &       & q^{-10}  &       &      &   & \\
                            &    q^{6}     &       &      &       &      &   & \\
                            &          &       &      &       &  q^{-9}  &   & \\
                            &          &    q^{3}  &      &       &      &   & \\ \hline
                       q^{6}    &          &       &      &       &      &   & \\
                            &          &       &      &  1    &      &   & \\
                            &          &       &      &       &      & q^{3}  & \\
                            &          &       &      &       &      &   & 1\\
\end{array} 
\right]
\\
%%%%%%%%%%%%%%%%%%%%%%%%%%%%%%%%%%%%%%%%%%%%%%%%%%%%%%%%%%%%%%%%%%%%%%%%%%%%%%%%%%%%%%%
\end{array}\]
%%%%%%%%%%%%%%%%%%%%%%%%%%%%%%%%%%%%%%%%%%%%%%%%%%%%%%%%%%%%%%%%%%%%%%%%%%%%%%%%%%%%%%%%%
\\
\np
\noi
{\bf Part C. Summary; Follow-up}

The methods presented above provide a means for constructing folds and characterizing
disjoint topological spaces defined by each distinct knot. In a following paper we
will further discuss the properties of these subspaces and how they might be used
in the partition function for statistical purposes.
\\
\\

The broader implications of this work were not apparent at first.
Initially,
I had thought of the $[T]$ product in the context of polynomial knot invariants,
guided by an analogy with the Artin/Burau group representation. Only afterward 
did it appear that the $[T]$ product was not unlike the Wilson Loop which is
used in quantum field theory: this expression could be used to derive
the Jones polynomial (Witten, Kauffman, Baez and Muniain).
However,
the matrices appear to contain more information than the polynomials;
for example, subspaces apparent in the matrix suggest symmetries
among subsets of crossings and might act to better characterize knot
classes. Furthermore, close relation, at least in appearance,
to the partition function suggests additional associations.

And yet, the crossings are, after all, not physical. The internal symmetries
of knots have long been a matter of   considerable interest
(Hoste et al, Gruenbaum and Shepard) and some of the information
pertaining to these symmetries is apparent in the mixing (or non-mixing)
of crossings as indicated in the knot matrices. For example, one would
expect knots $3_{1},5_{1},7_{1},...$ to be diagonal, as they are (see below). 

The degree of mixing of the crossings, as basis vectors, is in some
sense a measure of the complexity of the knot. The entropy involved in
such configurations has recently received considerable attention
(Baiesi et al).

Some open questions to be considered in the follow-up paper:\\
\noi
1. it is not clear how to relate the matrices to invariant polynomials,
such as Alexander/Jones/Kauffman, nor how, if at all, to construct
skein relations for the $[T]$ products\\
\noi
2. one change in methodology, possibly, periodic boundary conditions suggest that 
the $\tau$ matrix should, for the last crossing on the path be
\[
\left[
\begin{array}{ccc}
                        0  &   ...    &     e^{i\af t} \\
                           &          &            \\
                        1   &   ...    &          0 \\
\end{array} 
\right]
\]
instead of
\[
\left[
\begin{array}{ccc}
                        I_{N-1}  &   ...    &      \\
                                 &    0      &   e^{i\af t} \\
                                 &    1     &       0 \\
\end{array} 
\right]
\]
This change in procedure would assure the diagonal behavior of the simple
odd numbered knots.\\
\noi
3. the notion of curvature of a knot is admittedly vague, unlike that of, say
a 3-d manifold
\np
\noi
\ls{\bf References:}\ns

In getting started on this topic, the references that I found most
useful were: K Freed, {\em Renormalization Group Theory of Macromolecules};
C Livinston {\em Knots}; L Kauffman {\em Knots and Physics};
J Baez and J Muniain {\em Gauge Theories, Knots and Gravity}
\\
\\
\noi
\ls{\bf References to Part A:}\ns\\
\noi
A Ashtekar,  ed. A. Ashteker {\em Conceptual Problems of Quantum Gravity} (1991)
 Birkhauser Boston\\
\noi
WR Taylor;{\em Protein Knots and Fold Complexity};
Comp. Biol. Chem.  \underline{31} 151 (2007)\\
\noi
V Pande A Y Grosberg, T Tanaka;
 {\em Statistical Mechanics of Simple Models of Protein Folding and Design};
  Biophysical Journal \underline{73} 3192 (1997)\\
\noi 
N Madras  {\em Self-Avoiding Walk} Birkhauser (1993)\\
\noi
D Amit G Parisi L Peleti;
{\em Asymptotic Behavior of the True Self-Avoiding Random Walk};
Phys Rev B \underline{27} 1635 (1983)\\
\noi 
K.Freed {\em Renormaliztion Group Theory of Macromolecules} Wiley (1987)\\
\noi
JD Bryngelson;
{\em When is a Potential Accurate Enough for Structure Prediction};
 J Chem Phys \underline{100} 6038 (1994)\\
\noi
KA Dill, HS Chan;
{\em From Levinthal to Pathways to Funnels};
Nature Structural Biology \underline{4} 10 (1997)\\
\noi 
MA Moret MC Santana GF Zebende PG Pascutti;
{\em Self-similarity and Protein Compactness};
Phys Rev E \underline{80} 041908 (2009)\\
\noi 
JS Balafas TG Dewey;
{\em Multifractal Analysis of Solvent Accessibilities in Proteins};
Phys Rev E \underline{52} 880 (1995)\\
\noi
M Karplus;
{\em The Levinthal Paradox};
Folding and Design \underline{2} 569 (1997) \\
\noi
\\
\noi
\ls{\bf References to Part B}\ns\\

\noi
C Rovelli in {\em Knots Topology Quantum Field Theory} ed. L Lusanna World Scientific (1989)
\noi
L Kauffman; {\em Knots and Physics}; World Scientific (1992)\\
\noi
D Bolinger, Sulkowska J,;
{\em A Stevedore's Protein Knot};
PLoS Comp. Bio. \underline{6} e1000731 (2010)\\
\noi
D Meluzzi, Smith DE, Arya G,;
{\em Biophysics of Knotting};
Ann. Rev. Bioph. \underline{39} 349 (2010)\\
\noi
S Wallin, Zeldovich KB, Shakhnovich EI;
{\em Folding Mechanics of a Knotted Protein};
J Mol. Bio. \underline{368} 884 (2007)\\
\noi
C Livingston ; {\em Knot Theory} ; Math Assoc Amer Pub (1993)\\
\noi
J S Birman  T E Brendle {\em Handbook of Knot Theory}
W Menasco M Thistlethwaite; Elsevier (2005)\\
\noi
J Hoste, M Thistlethwaite, J Weeks 
{\em The First 1,701,936 Knots};
Math. Intel. \underline{20} 33 (1998)\\
\noi
F Y Wu;
{\em Knot Theory and Statistical Mechanics};
Rev Mod Phys \underline{64} 1099 (1992)\\
\noi
V Manturov {\em Knot Theory} Chapman Hall CRC (2004)\\
\noi
M Baiesi, E Orlandini, AL Stella;
{\em The Entropy Cost to Tie a Knot};
 J Stat. Mech.  (2010) P06012;
 arxiv:1003.5134v1 cond-mat.stat-mech\\
\noi
J Baez J Muniain {\em Gauge Theories, Knots and Gravity} World Scientific (1994)\\
\noi
E Witten;
{\em Quantum Field Theory and the Jones Polynomial};
Comm Math Phys \underline{121} 351 (1989)\\
\noi
 B Gruenbaum GC Shephard;{\em Symmetry Groups of Knots};
Math Mag \underline{58} 161 (1985)
\\
\noi
{\bf Acknowledgements:} I wish to thank the physics department at
UW-Madison for a fellowship during which this work was begun.

\np
%%%%%%%%%%%%%%%%%%%%%%%%%%%%%%%%%%%%%%%%%%%%%%%%%%%%%%%%%%%%%%%%%%%%%%%%%%%%%%%%%%%%%%%%%%%%%%%%%5
{\bf APPENDIX}\\
\\
\begin{figure}[!h]
\begin{center}
\includegraphics[width=3in,height=3in,trim=1in 2in 1in 2in, angle=-90,scale=1.2]{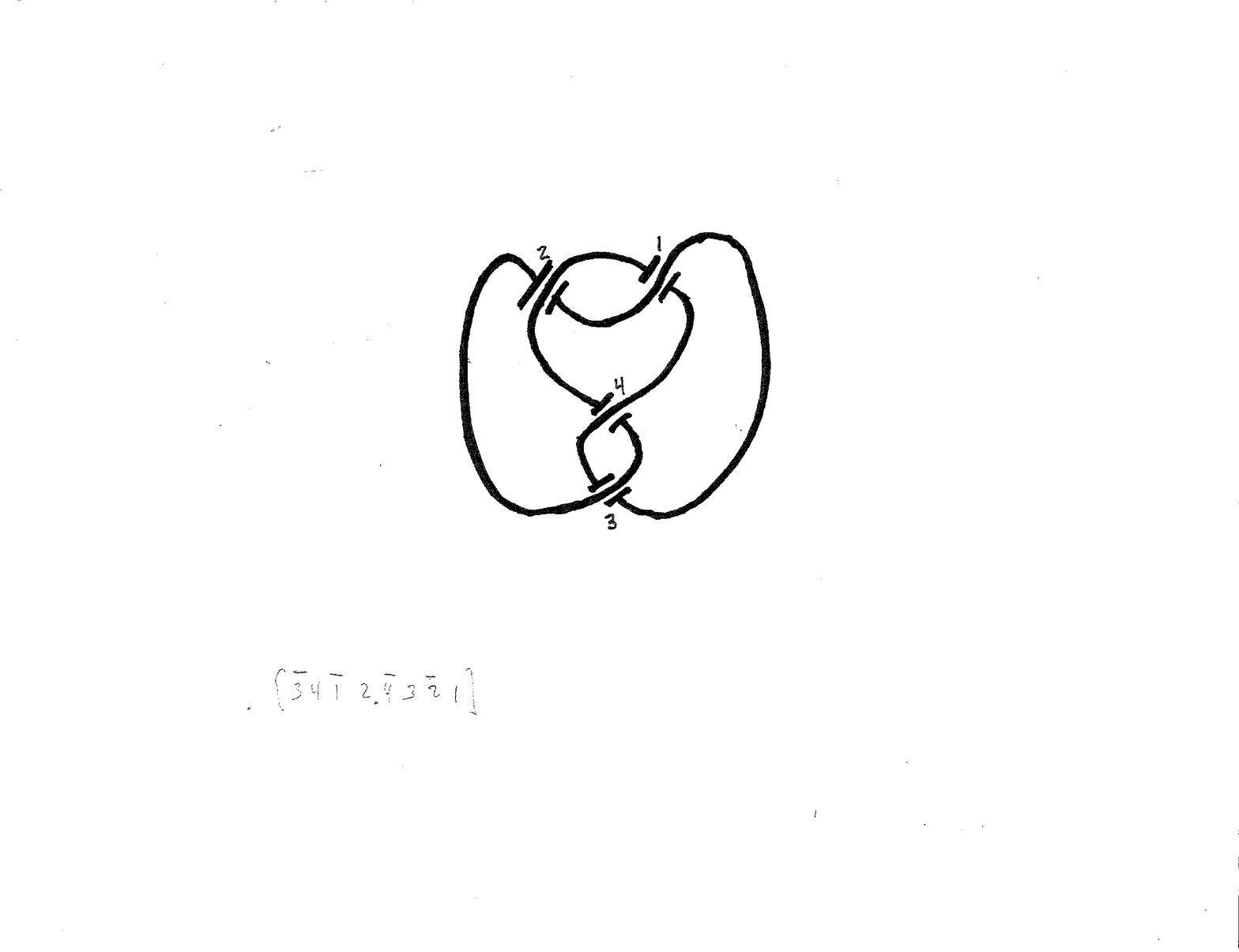}
\caption{ $4_{1}$ knot }
\end{center}
\end{figure}

\begin{figure}[!h]
\begin{center}
\includegraphics[width=3in,height=3in,trim=1in 1in 1in 1in,scale=1.1]{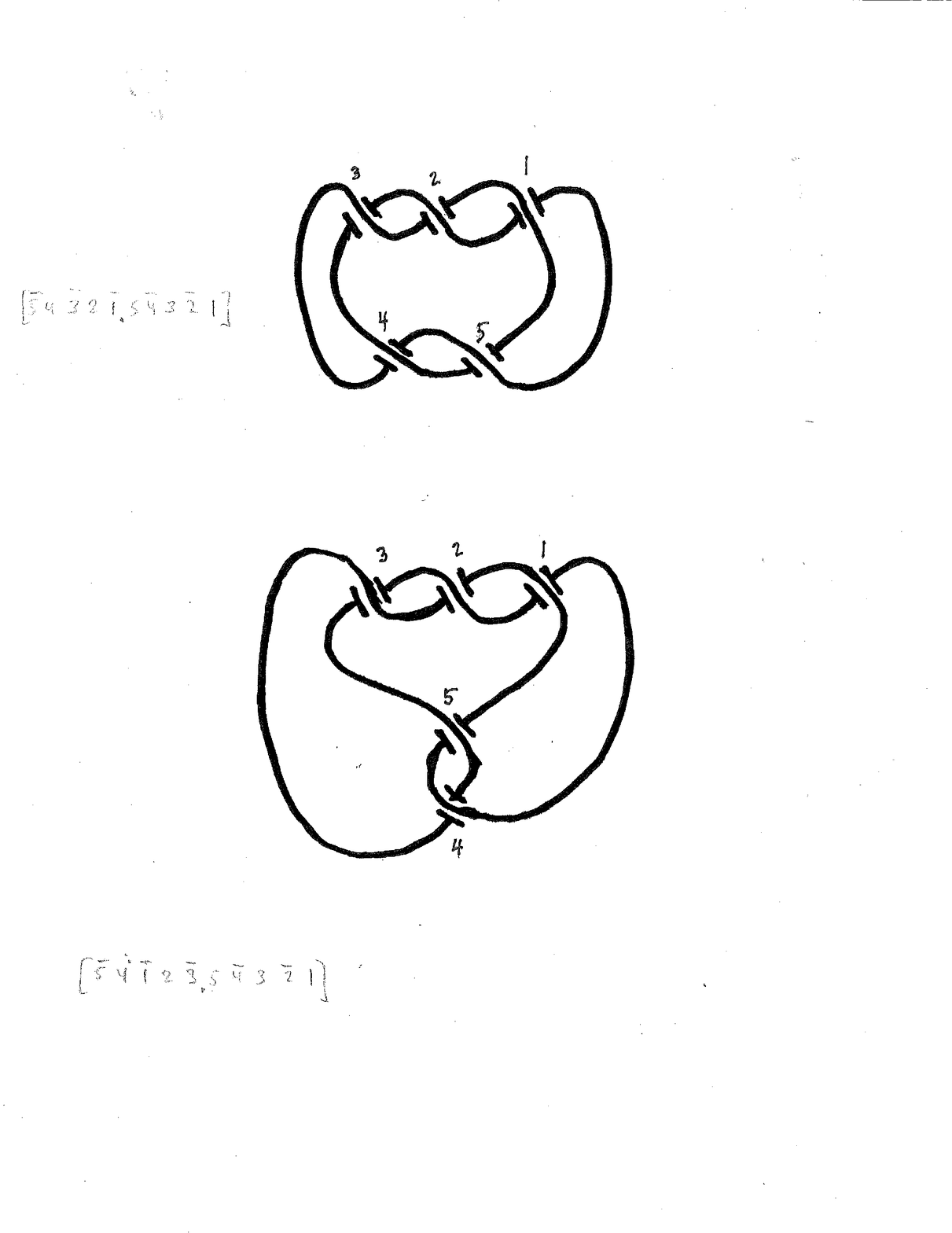}
\caption{ $5_{1}$ $5_{2}$ knots }
\end{center}
\end{figure}

\begin{figure}[!h]
\begin{center}
\includegraphics[width=3in,height=3in,trim=1in 1in 1in 1in,scale=1.1]{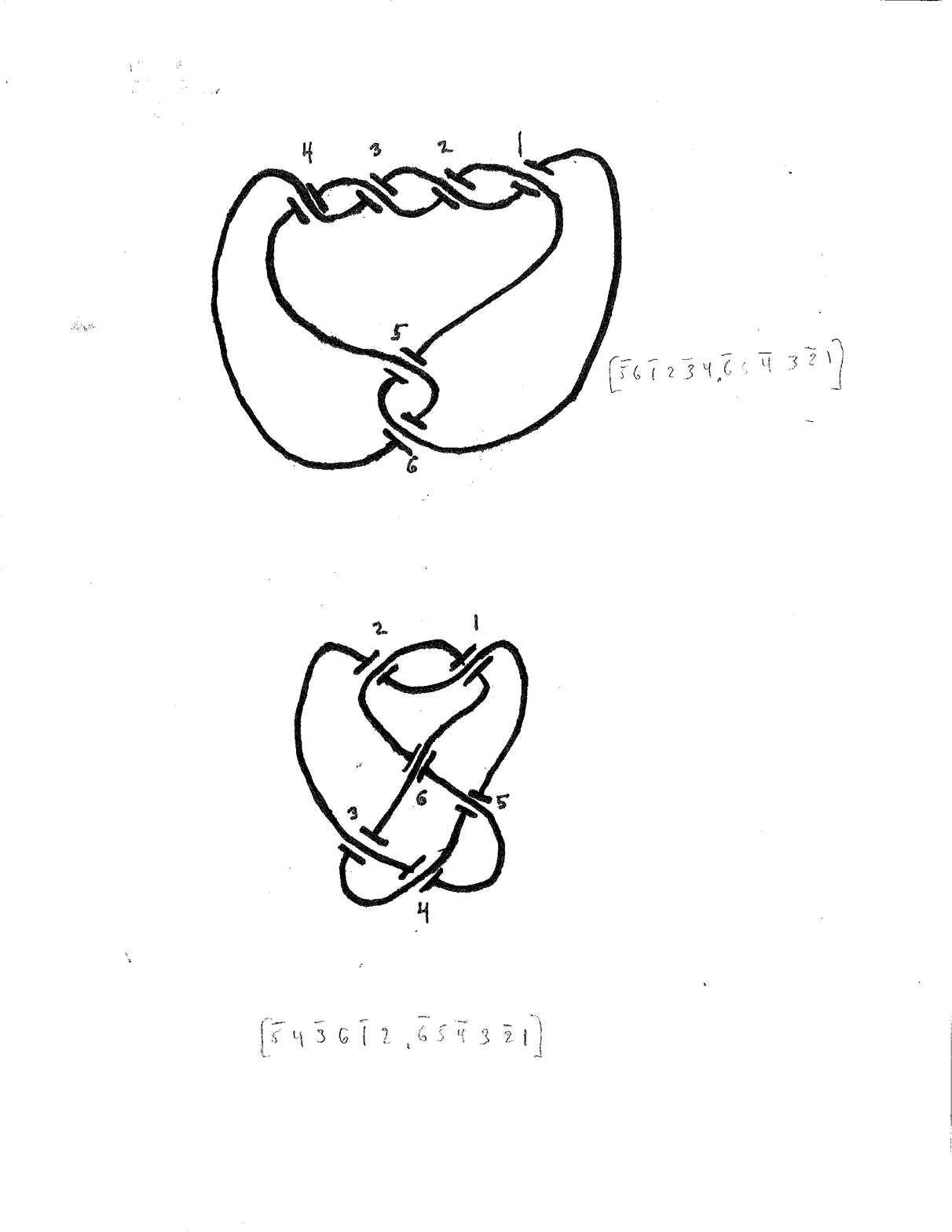}
\caption{ $6_{1}$ $6_{2}$ knots}
\end{center}
\end{figure}

\begin{figure}[!h]
\begin{center}
\includegraphics[width=3in,height=3in,trim=1in 2in 1in 2in,scale=1.1] {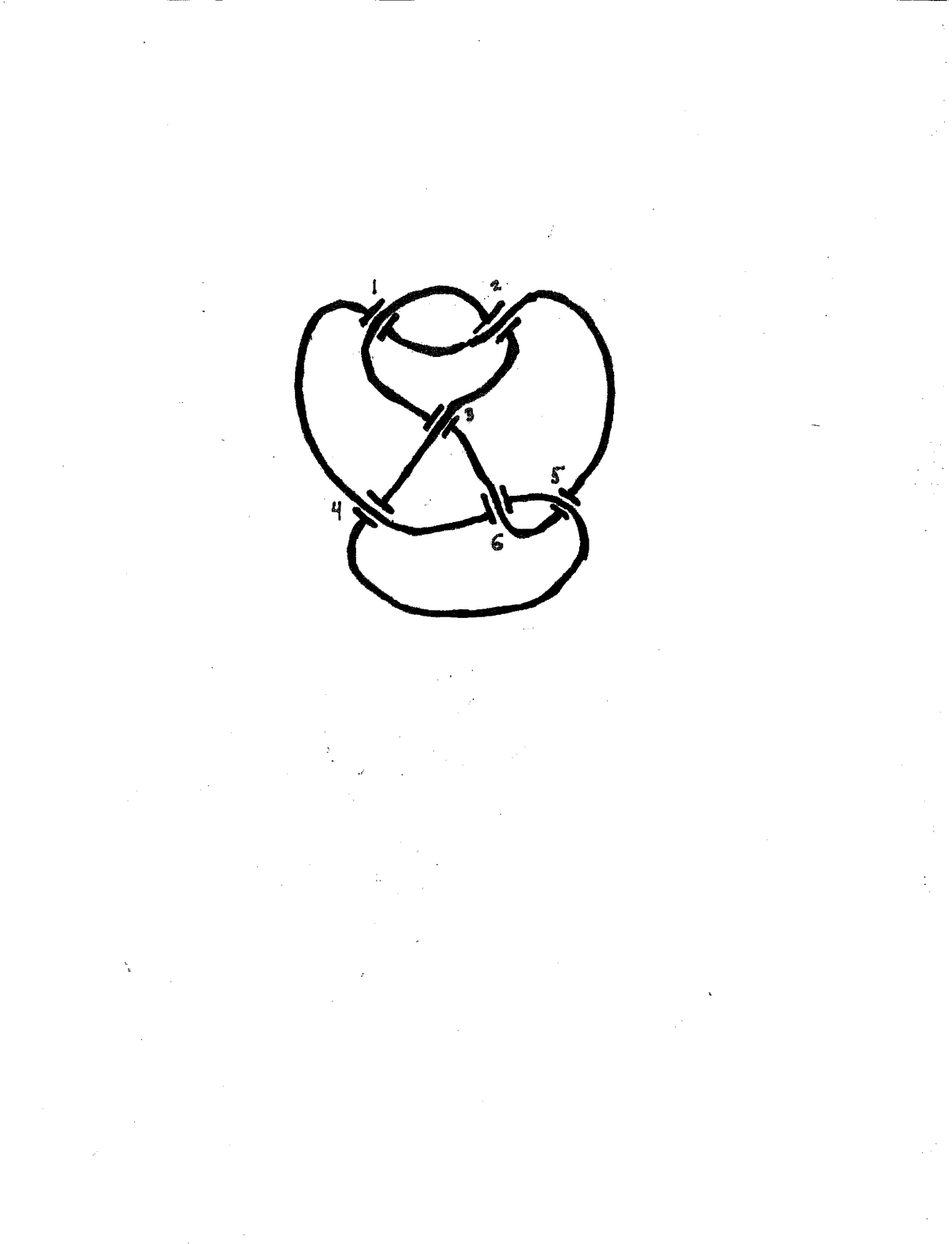}
\caption{ $6_{3}$ knot}
\end{center}
\end{figure}

\begin{figure}[!h]
\begin{center}
\includegraphics[width=5in,height=5in]{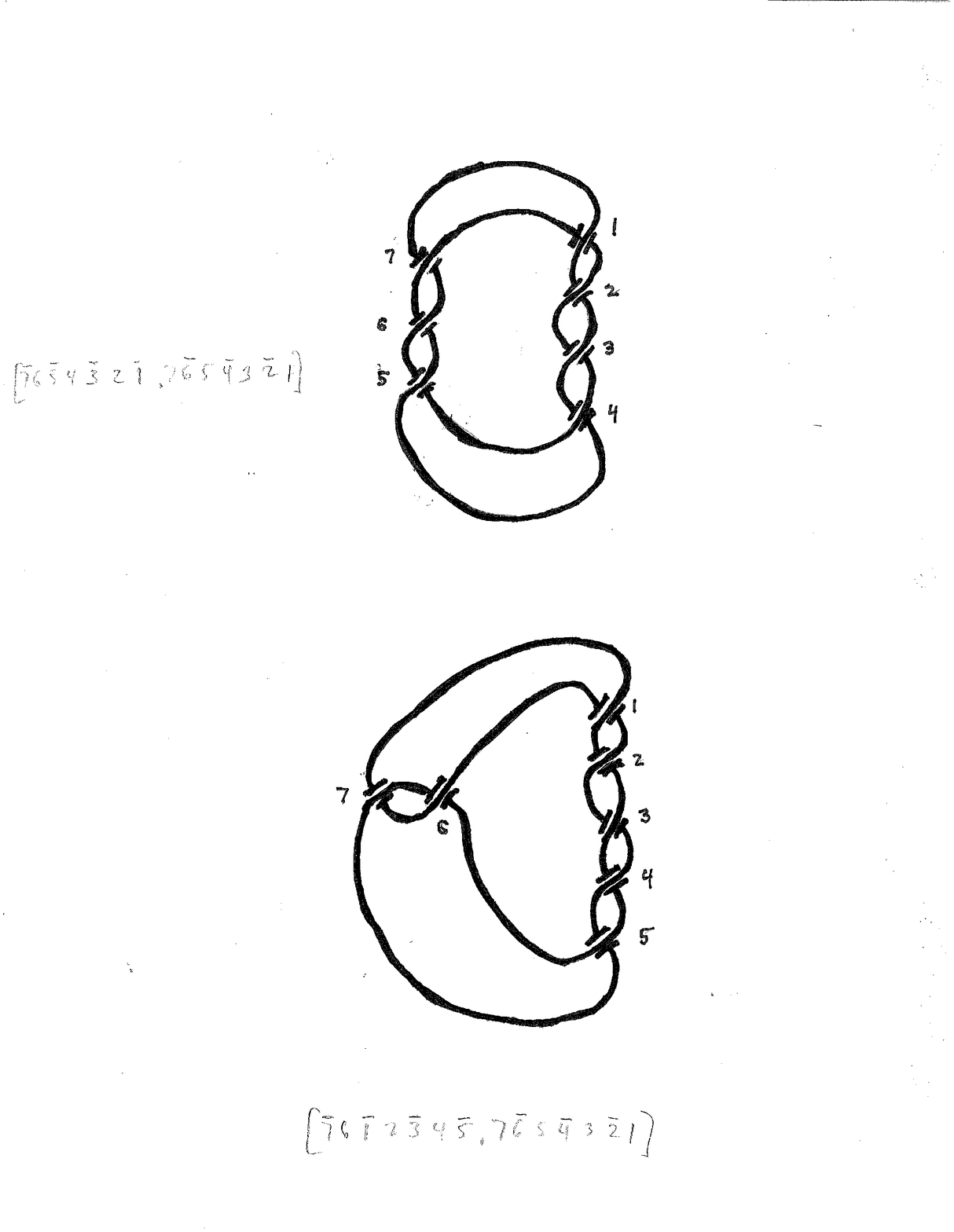}
\caption{ $7_{1}$ $7_{2}$}
\end{center}
\end{figure}

\begin{figure}[!h]
\begin{center}
\includegraphics[width=5in,height=5in]{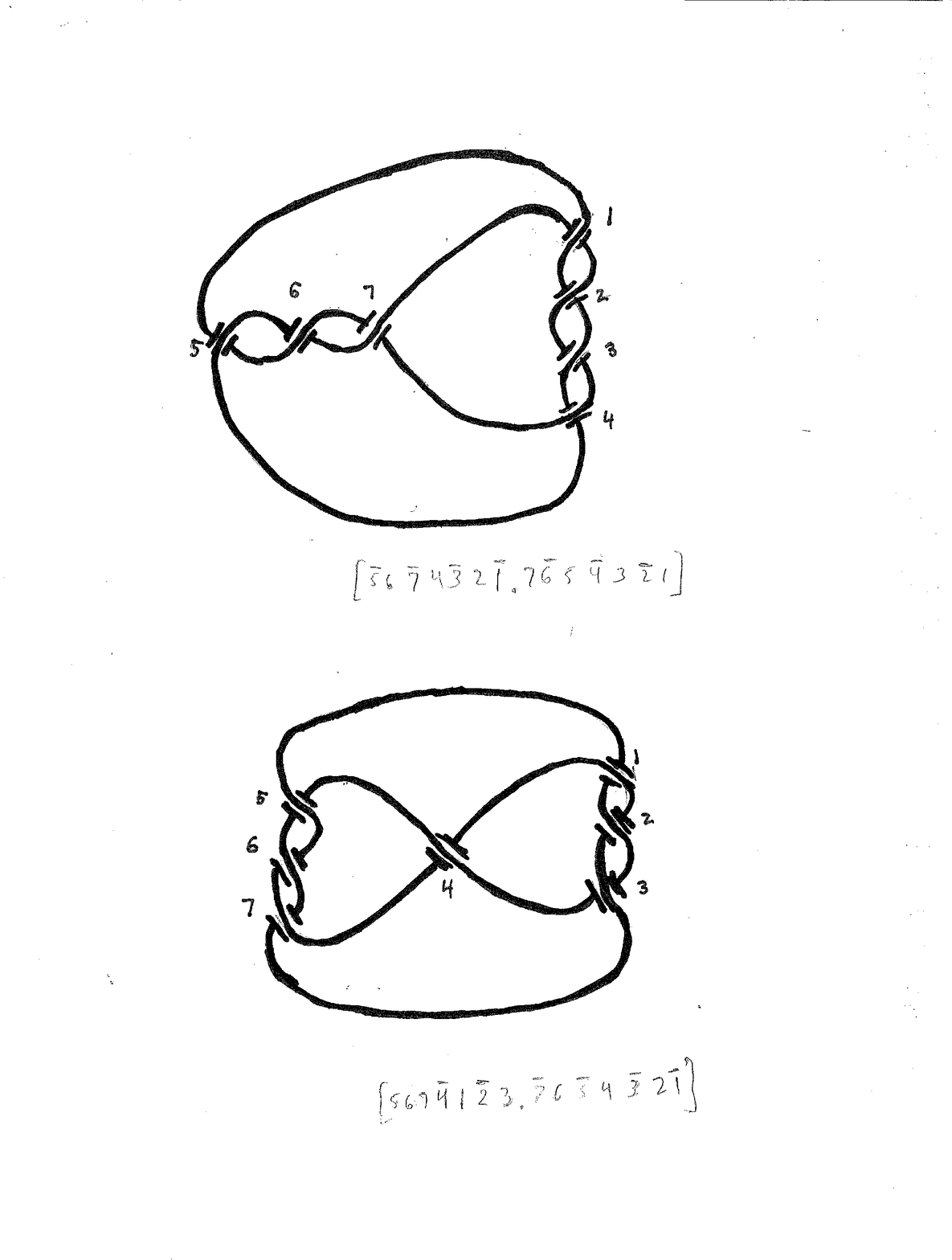}
\caption{ $7_{3}$ $7_{4}$}
\end{center}
\end{figure}

\begin{figure}[!h]
\begin{center}
\includegraphics[width=5in,height=5in]{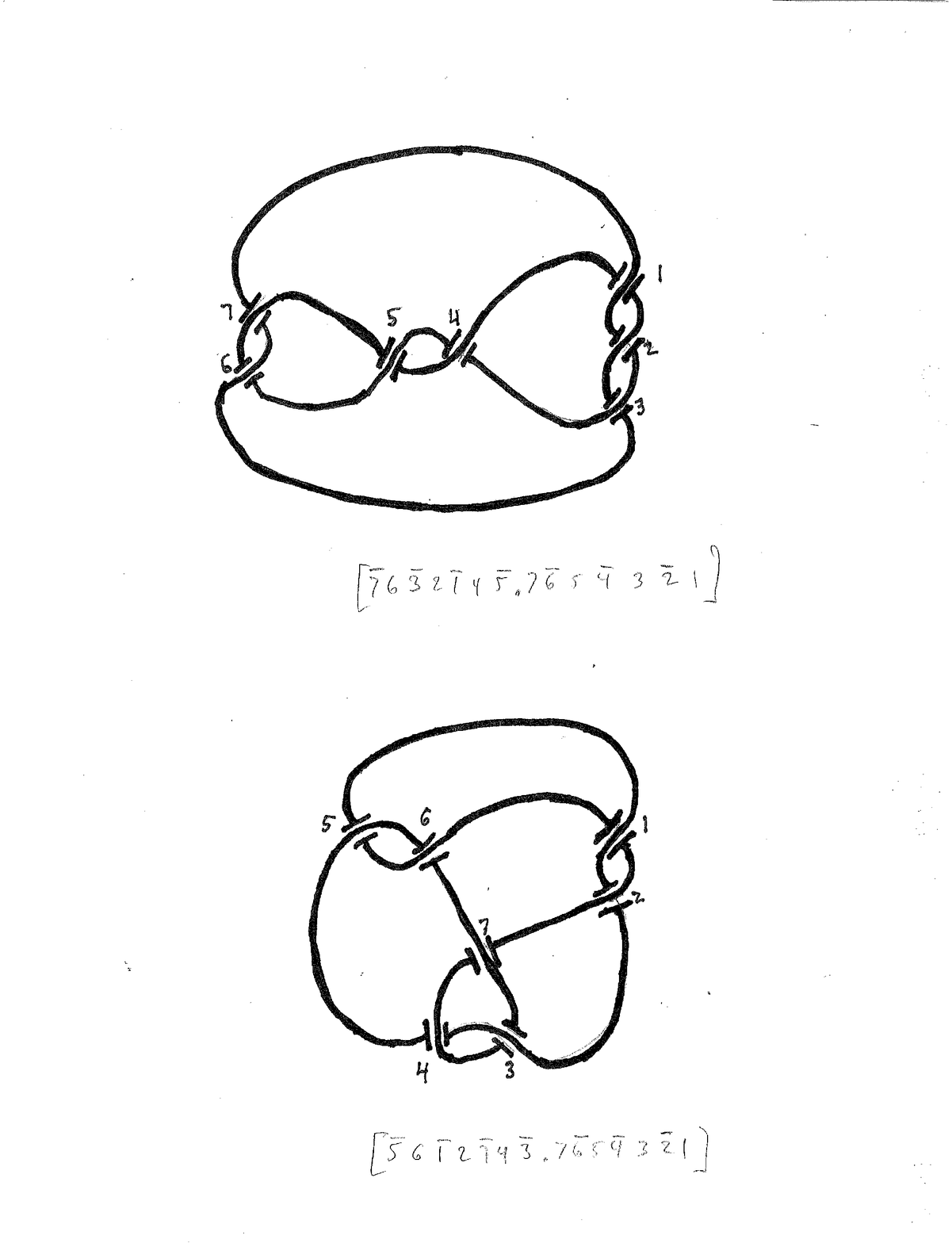}
\caption{ $7_{5}$ $7_{6}$}
\end{center}
\end{figure}

\begin{figure}[!h]
\begin{center}
\includegraphics[width=5in,height=5in]{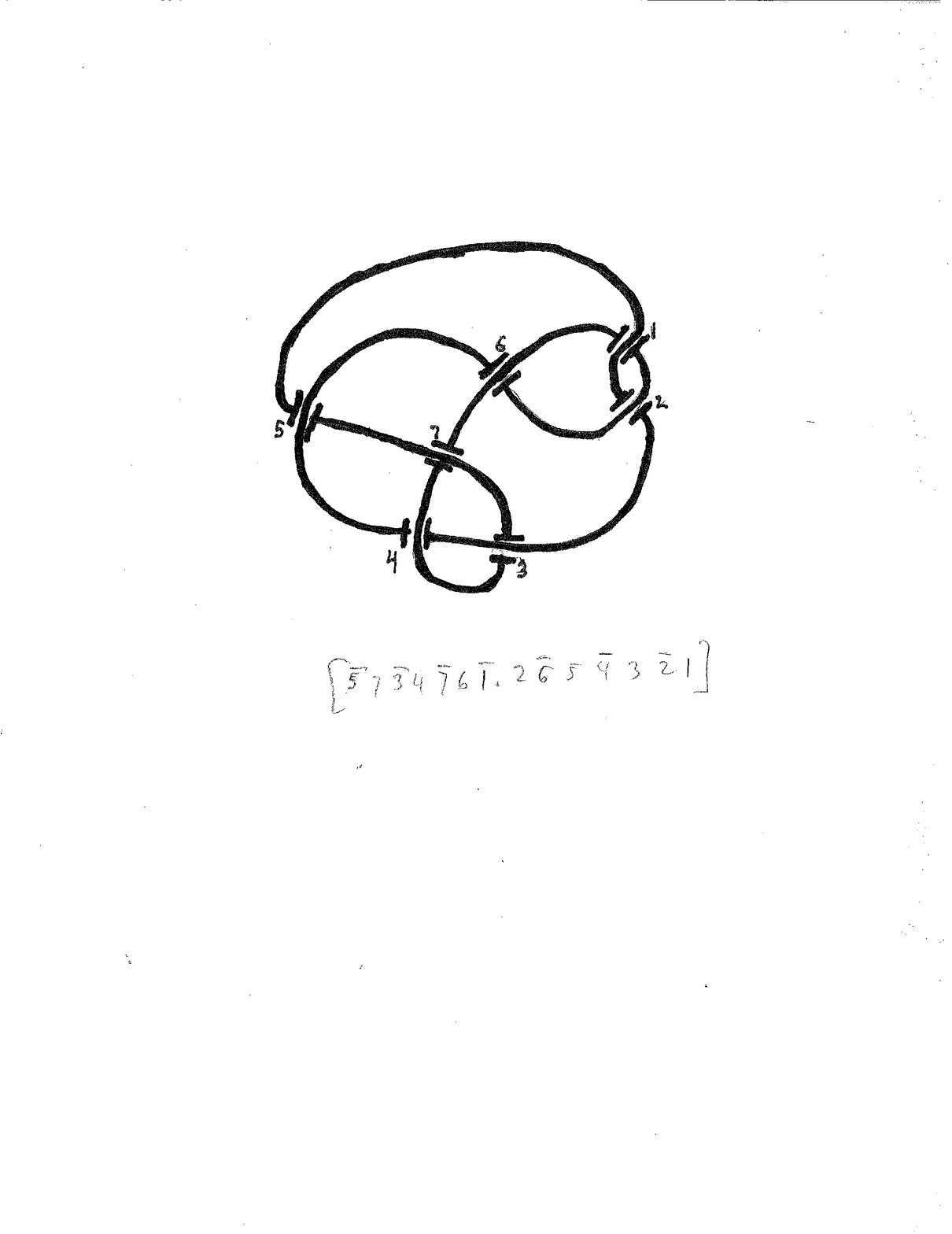}
\caption{ $7_{7}$}
\end{center}
\end{figure}
%%%%%%%%%%%%%%%%%%%%%%%%%%%%%%%%%%%%%%%%%%%%%%%%%%%%%%%%%%%%%%%%%%%%%%%%%%%%%%%%%%%%%%%%%%%%%%

\end{document}